\documentclass[twocolumn,aps, pra, nofootinbib,superscriptaddress,showkeys,showpacs,longbibliography]{revtex4-2}
\usepackage{amsmath,amsfonts,amssymb,amstext,amscd,amsthm,dsfont}
\usepackage{physics}
\usepackage{color}
\usepackage[colorlinks, citecolor=blue]{hyperref}
\usepackage{amsfonts}
\usepackage[font=small,format=plain,labelsep=period,labelfont=bf,textfont=normal,singlelinecheck=false,justification= raggedright]{caption}
\usepackage{braket}
\usepackage{graphicx}
\usepackage{subcaption}
\usepackage{times}
\usepackage{color}
\usepackage{soul,xcolor}
\usepackage{bm}
\usepackage[normalem]{ulem}
\usepackage{ragged2e}

\newcommand{\prague}{FNSPE, Czech Technical University in Prague, Br\^ehov\'{a} 7, 119 15, Praha 1, Czech Republic}

\begin{document}

\title{Quantum metrology in a nonlinear-interferometer with feedback}

\author{Shivani Singh}
\email{singhshi@fjfi.cvut.cz}
\affiliation{%
\prague} 
\author{Craig S. Hamilton}
\email{hamilcra@fjfi.cvut.cz}
\affiliation{\prague}
\author{Igor Jex}
\email{igor.jex@fjfi.cvut.cz}
\affiliation{\prague}

\begin{abstract}

In this paper, we propose a nonlinear interferometer with feedback loops and explore its efficiency for phase estimation. 
We analyse two feedback schemes, one where both modes of the interferometer are fed-back into the device and another where only one mode is fed-back. 
The quantum Fisher information (QFI) for phase estimation in each feedback scheme increases with each feedback loop, and similar to the standard SU(1,1) nonlinear interferometer, phase estimation in this scheme is sensitive to photon loss when the inputs are vacuum state.
In terms of resources, we show that, in the low-loss regime, our scheme performs better than standard nonlinear interferometer.
The feedback scheme provides the minimum phase variance when the unknown phase is small. 
We have also provided a special case where feedback scheme provided enhanced QFI even for large phase values. This is achieved by switching between squeezing and anti-squeezing operators after every few loops.

\end{abstract}

\maketitle
\vskip -0.7in
\noindent


\section{Introduction}

Quantum phase estimation is an integral part of quantum sensing, imaging, metrology, and many communication protocols\,\cite{CFP17,M16,MK93,GL06}.
Due to immense demand on precise phase estimation, quantum metrology explores the possibility of using quantum resources to enhance the phase sensitivity beyond classical limits\,\cite{DD09,GO12,DZ20,AP97,SB03}.
Phase sensitivity in traditional interferometers, using classical resources, is limited by the Shot Noise Limit (SNL) that scales as $1/\sqrt{N}$ but an interferometer with quantum resources can beat SNL and can even achieve the Heisenberg limit (HL), which scales as $ 1/N$, where $N$ is the average photon number (or intensity) that probes the unknown phase. 
Therefore, considerable effort has been put into enhancing the phase sensitivity in interferometers by employing different quantum resources such as input states, various dynamics, and measurement strategies\,\cite{AA10,NO07,AR10,D08,JM11,CGB03}. 
Nonlinear interferometers are one such configuration that have been used to enhance phase sensitivity beyond the standard quantum limit (SNL)\,\cite{YM96,AG17,LZ23}. They were first proposed by Yurke et al.\,\cite{YM96}, who replaced the operation of a Mach-Zehnder interferometer — belonging to the SU(2) group — with one based on the SU(1,1) group, and thus called it an SU(1,1) interferometer.
In this configuration, the beam splitters of the Mach-Zehnder interferometer are replaced with optical parametric amplifiers (OPAs) or four-wave mixers, which generate a two-mode squeezed state that is entangled. Phase estimation inside the SU(1,1) interferometer is improved due to the presence of the entangled probe state within the interferometer. This idea has led to the development of various nonlinear interferometer schemes \,\cite{C20,CZ22,SL17,DC20}.


The standard SU(1,1) interferometer consists of two OPAs, where the first OPA generates an entangled state that probes the unknown phase in the system and a second OPA that is the inverse transformation of the first OPA process.  
The phase sensitivity in the standard SU(1,1) interferometer is below the SNL and Marino et.al.\,\cite{MC12} show that it is sensitive against photon losses when the initial state is the vacuum state.  
Another variant of the nonlinear interferometer is based upon two-port feedback optical parametric amplification (NITFPA) that consist of a nondegenerate optical parametric amplifier and two linear beam splitters and has been used for entanglement enhancement in the nonlinear interferometer\,\cite{JX20} as well as for precision enhancement\,\cite{HZ21,G24,LX21}.

In this work, we present a theoretical protocol that uses the SU(1,1) interferometer in feedback loop where the output
of the SU(1,1) interferometer is fed-back into the set-up. The
configuration of this setup is similar to sequential computation\,\cite{BG15, M11}. The continuous state feedback scheme that
is used here is a non-adaptive feedback scheme that has been
used in other quantum optical protocols\,\cite{JX20, W94, EL20, ZL17}.
It is a compact setup that recycles the resources in the system using the loop.  The setup of our SU(1,1) interferometer with feedback  is analogues to driven discrete-time quantum walk\,\cite{CI17}, where similar to the feedback scheme the final state is given by the unitary operation followed by an intricate squeezing operation, and studies show that the driven quantum walk provides enhanced precision in parameter estimation \,\cite{SC23}.

The quantum Fisher information (QFI) quantifies the precision of an estimation of an unknown parameter, in our case a relative phase shift between two modes of the interferometer.  
It is the intrinsic information of an unknown parameter in the quantum state and thus it is not associated with the measurement technique\,\cite{H11,P09,TA14,GL04}. According to the quantum Cramer-Rao theorem, the variance of an estimated parameter is lower bounded by the QFI.
In this work we have focused on devising a protocol to enhance the phase estimation using SU(1,1) interferometer-based configurations, analogues to sequential computation schemes\,\cite{BG15, M11}. We show that the parameter estimation in this scheme is improved when compared to the standard SU(1,1) interferometer. 

\begin{figure}
\centering
\includegraphics[width=0.49\textwidth]{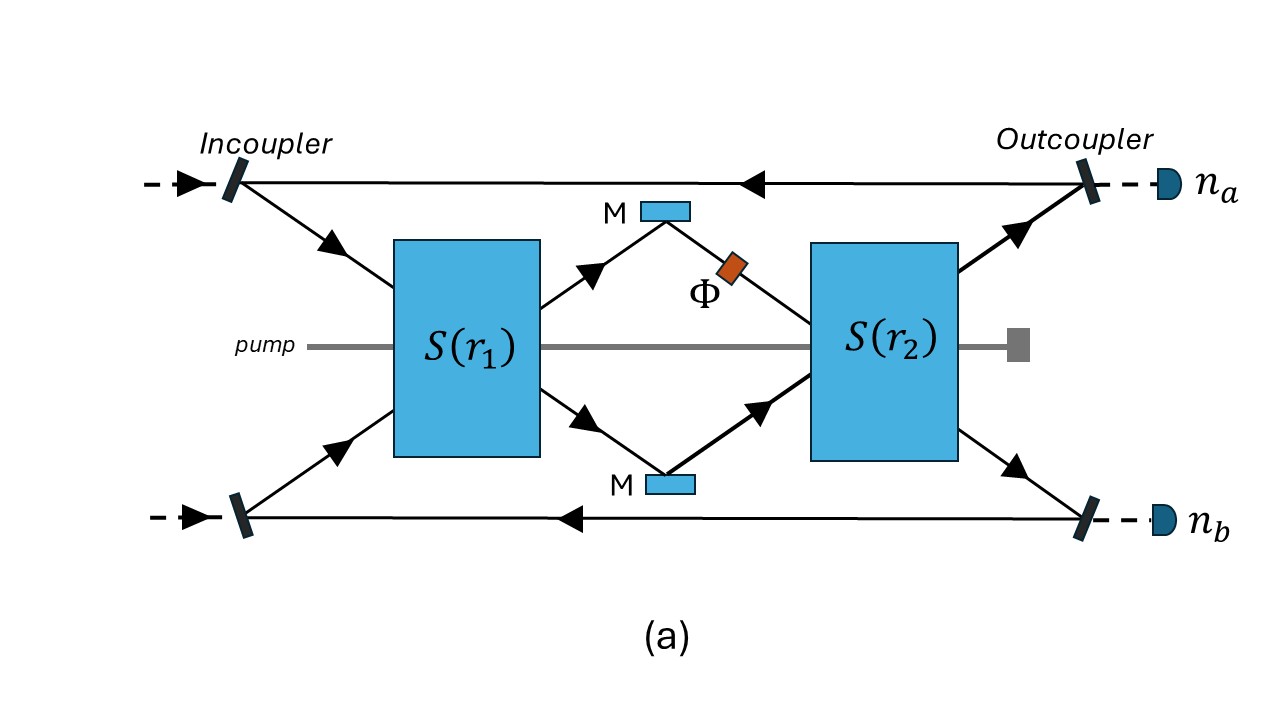}
\includegraphics[width=0.49\textwidth]{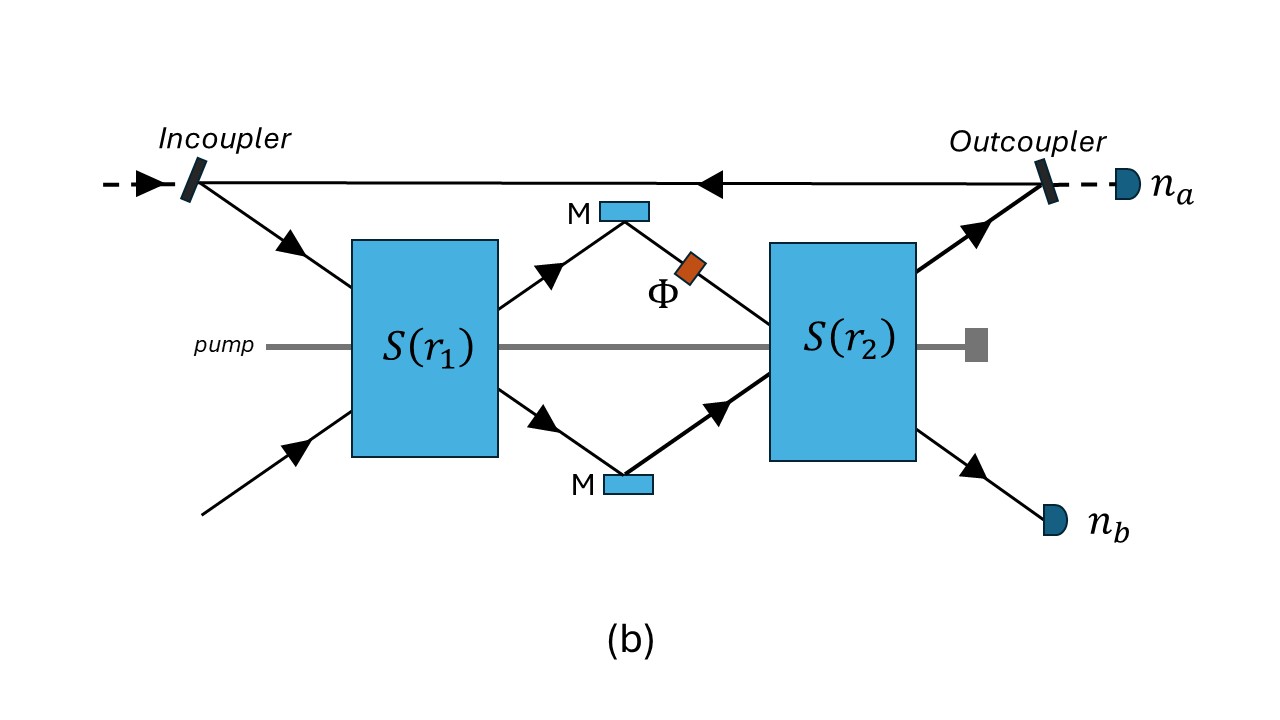}
\caption{Illustration of SU(1,1) interferometer (a) with feedback in both the modes, called sequential feedback scheme, and (b)  with feedback in only one arm, called partial feedback scheme. Initial states are vacuum state, $\Phi$ is the unknown phase operator, and $S(r_1)$ and $S(r_2)$ are the two-mode squeezing operators.  }
\label{SU11-a}
\end{figure}

Our first scheme, which we term the sequential feedback scheme, is when both the modes of the SU(1,1) interferometer are in the feedback loop, as shown in fig.\,\ref{SU11-a}-(a). Measurement is performed after $N$-loops. 
Our study shows that the photon intensity in this scheme oscillates between a maximum and minimum value and the interval of oscillation depends upon the unknown phase. The QFI in this scheme also depends upon the phase to be estimated and increases quadratically with the number of loops.
In this feedback scheme, the enhancement in the QFI is mainly due to the multiple interactions with the phase shifter. We have also devised a partial feedback scheme in which the QFI increases monotonically due to the the enhanced nonclassicality of the state.
The second scheme, which we term the partial feedback scheme, is when only one mode of the SU(1,1) interferometer is in the loop while the other mode is measured as shown in fig.\,\ref{SU11-a}-(b).
In this case, photon intensity and QFI, both increase with the number of loops, throughout the process. Here also, the QFI depends upon the unknown phase. The QFI in both the feedback schemes arises from the underlying dynamics, where both multiple passes through the phase shifter and the nonclassicality of the generated state play significant roles.

The paper is organized as follows- section\,\ref{SU11_sNa} describes the SU(1,1) interferometer with feedback loop and also discusses the phase-space formalism of it. 
In section\,\ref{secQFI}, we have analysed the QFI in our feedback schemes of the SU(1,1) interferometer, that changes with each iteration due to the change in the photon number in the interferometer that probes the phase.
In the subsection\,\ref{secNoisy} we have studied the QFI in the presence of photon loss in SU(1,1) interferometer with feedback (noisy feedback scheme) and subsection\,\ref{secSensitivity}, the phase sensitivity of the feedback schemes and compared it with the standard SU(1,1) interferometer. 
In section\,\ref{secSpecial}, we discuss a special case of the sequential scheme, where the squeezing operation is swapped after a fixed number of loops in order to estimate the small change in the initial phase more efficiently. In the last section\,\ref{concl}, we have concluded the outcomes of the analysis with some discussion.


\section{SU(1,1) interferometer with feedback loop } \label{SU11_sNa}

\begin{figure}
\centering
\includegraphics[width=0.49\textwidth]{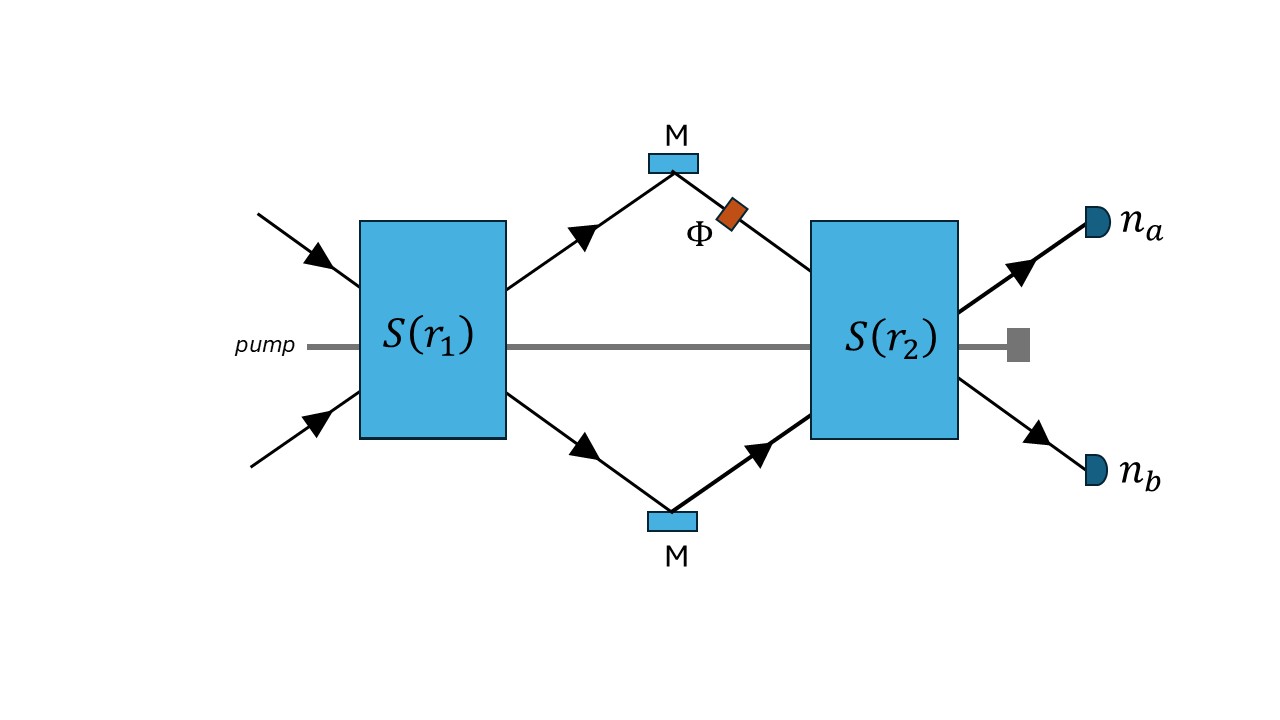}
\caption{An illustration of SU(1,1) interferometer where $\Phi$ is the unknown phase operator, and the two 
-mode squeezing operators are  $S(\Gamma_1) = S(r_1)$ and $S(\Gamma_2) = S(r_2)$. The initial states are the vacuum states. }
\label{SU11}
\end{figure}

The SU(1,1) interferometer, shown in Fig.\,\ref{SU11},  consists of a pair of two-mode squeezing operations with the unknown phase, $\phi$, in one mode. The input state to the interferometer is the vacuum state, which evolves to two mode squeezed state under first squeezing operator $S(\Gamma_1) = \exp\{\Gamma_1(\hat{a}\hat{b}) - \Gamma_{1}^{*}(\hat{a}^{\dag}\hat{b}^{\dag})\}$, with squeezing parameter  $\Gamma_1 = r_1 e^{i\theta_1}$. 
One mode of the state then acquires the unknown phase $\phi$ (through the unitary operator $U(\phi) = \exp\big(i \phi (\hat{a}^{\dag}\hat{a} - \hat{b}^{\dag}\hat{b})/2 \big)$) that needs to be estimated. 
Finally, another squeezing operator $S(\Gamma_2)$ (where $\Gamma_2 = r_2 e^{i\theta_2}$) is applied before the detection. 
The output state of the SU(1,1)-interferometer can be written as,
\begin{align}\label{SUevo}
\ket{\Psi(\phi)} &= W \ket{0} \nonumber \\
&= S(\Gamma_2)U(\phi)S(\Gamma_1) \ket{0}
\end{align}
where, the two-mode squeezing operator is $S(\Gamma_{1(2)})$  and $U(\phi)$. 
The creation (annihilation) operators of the two modes are $\hat{a}^{\dag}$ ($\hat{a}$) and $\hat{b}^{\dag}$ ($\hat{b}$), respectively and $\ket{0}$ is the vacuum state.
In the standard nonlinear-interferometer $\Gamma_1=-\Gamma_2 = r$ and the intensity in each mode of it is given by $n = \langle \hat{a}^{\dag}\hat{a} \rangle = \sinh^{2}(r_1)$. 


\subsection{ SU(1,1) interferometer with feedback-}

When the output of a SU(1,1) interferometer is used to drive another SU(1,1) interferometer experiment, we define it as SU(1,1) interferometer with feedback.
The SU(1,1) interferometer configuration presented here consists of two-squeezing operators and input-output couplers as shown in Fig.\,\ref{SU11-a}.
We have analyzed two possible regimes of feedback, $(a)$ when both modes are fed back into the loop, which we term as sequential feedback scheme, and $(b)$ when only one mode of SU(1,1)-interferometer is fed-back while the other one is measured, which we term as the partial feedback scheme. 

\begin{widetext}

\begin{figure}[h]
\centering
\includegraphics[width=0.7\textwidth]{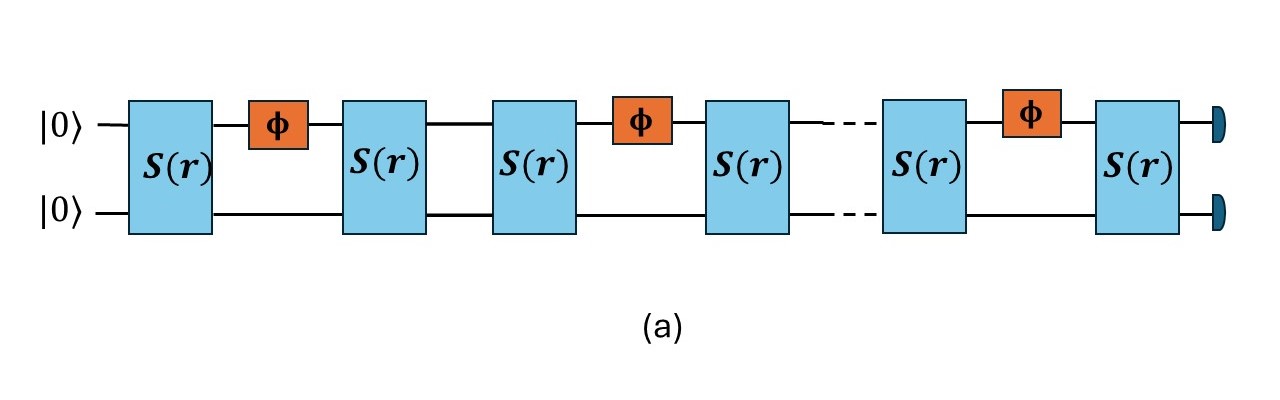}
\includegraphics[width=0.7\textwidth]{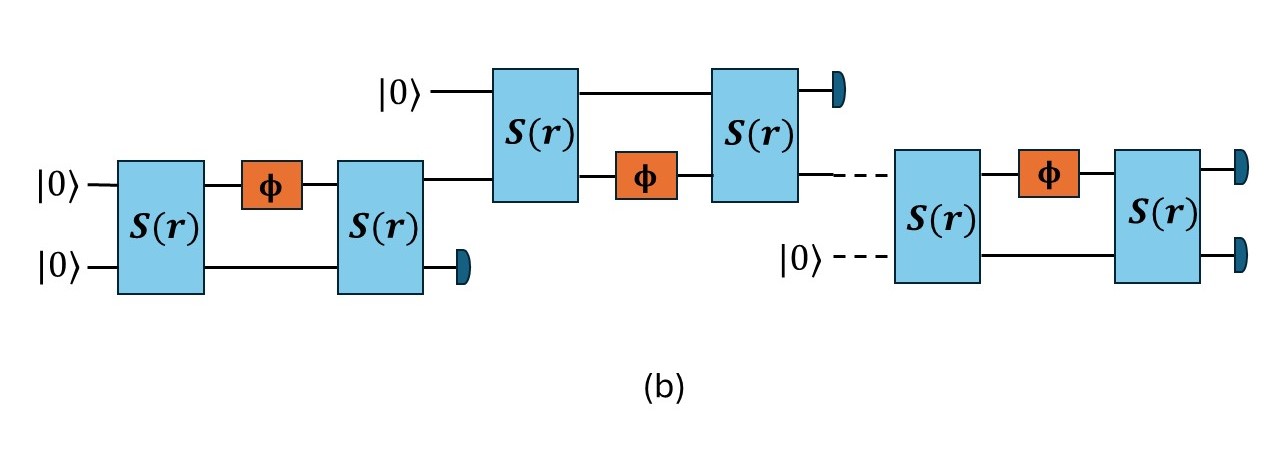}
\caption{Expansion of SU(1,1) interferometer in  (a) sequential feedback scheme, when both the ports of the interferometer are in feedback loop, and (b)  partial feedback scheme, when only one port is fed-back in the loop while other arm is reinitialized to vacuum state, respectively. The expanded form of the SU(1,1) interferometer  in partial feedback scheme can be mapped to $(N+1)$-modes where $N$ is the number of loops. Here, $\Phi$ is the unknown phase operator, and $S(r_1) = S(r_2) = S(r)$ is the two-mode squeezing operator. }
\label{SU11-b}
\end{figure}

\end{widetext}

{\it SU(1,1) interferometer in sequential feedback scheme -}
 The general setup of this scheme is shown in fig.\,\ref{SU11-a}-(a).
In this scheme, the input state goes through multiple iterations of the SU(1,1) interferometer without being reinitialized as shown in fig.\,\ref{SU11-b}-(a).  

The output of the SU(1,1)-interferometer in this scheme after $N$-feedback loops is given by,
\begin{align}\label{eqa2}
\ket{\Psi(\phi)}_{s} &= W_s \ket{0} \nonumber \\
&= \left[S(\Gamma_2)U(\phi)S(\Gamma_1)\right]^N \ket{0} \nonumber \\
\end{align}
where, $\Gamma_{1(2)} = r_{1(2)} e^{i\theta_{1(2)}}$. The evolution operator $ W_s$ is multiple applications of the evolution operator of the SU(1,1)-interferometer $W$ where $\Gamma_1 = \Gamma_2$, given in eq.\,\eqref{SUevo}.
Using the relation\,\cite{XW89},
\begin{align}\label{SUS}
S(\Gamma)e^{i \hat{a}^{\dag}\zeta\hat{a}} = e^{i \hat{a}^{\dag}\zeta \hat{a}} S(e^{-i\zeta}\Gamma e^{-i\zeta})
\end{align}
The output of the interferometer with $N$-feedback loops can be given as, 
\begin{align}\label{simple_loopeq}
W_s &= U(\phi)^N S(\Gamma_f(\phi))
\end{align}
where $S(\Gamma_f(\phi)) = S(e^{-iN\phi/2}\Gamma_2e^{-iN\phi/2})...S(\Gamma_1)$. Thus the state that probes the unknown phase embedded in $U$ is given by the squeezed state with an intricate squeezing parameter $\Gamma_f$ which is phase dependent. 

For simplicity, we will consider the squeezing parameter of both the squeezing operators in the evolution operator $W_s$ as same i.e., $\Gamma_{1(2)} = r$ such that the pump phase $\theta_{1(2)} = 0$ and the final squeezing is $S(\Gamma_f(\phi) = S(e^{-iN\phi/2}re^{-iN\phi/2})...S(r)$.  
This gives us the dependence of the final squeezing parameter $S(\Gamma_f)$ on the unknown phase $\phi$ and thus the intensity in terms of the $\phi$ as shown in the Fig.\,\ref{Int_sequential}.
It shows that the photon intensity oscillates with respect to the phase $\phi$ and the number of loops $N$. 
 The amplitude and interval of the oscillations decreases with the phase $\phi$. 
In this scheme, the modes interfere constructively due to the phase-matching condition initially, which enhances the intensity with the number of loops. However, after a certain number of iterations, the modes interfere destructively, since the phase in $S(\Gamma_f(\phi)$ starts to move away from the phase matched condition, and the intensity in the interferometer decreases. This process of phase matching and mismatching continues with the number of loops which leads to oscillating photon intensity in the sequential interferometer.

{\it SU(1,1)-interferometer in the partial feedback scheme -} 
The general setup of this scheme is shown in fig.\,\ref{SU11-a}-(b).
In this scheme only one mode of the SU(1,1)-interferometer is fed-back into the loop while the other mode is measured and then re-initialised in the vacuum state, as shown in Fig.\,\ref{SU11-b}-(b). It could also be re-initialised in another state to allow for adaptive processes, but in this we  focus only on the case where the input states are re-initialized  to the vacuum state.  

The output of this scheme, after $N$-feedback loops, is
\begin{align}\label{eqa1}
\ket{\Psi(\phi)}_{a} &= W_a \ket{0} \nonumber \\
&= S(\Gamma_2)U(\phi)S(\Gamma_1)\left[\hat{\Pi} S(\Gamma_2)U(\phi)S(\Gamma_1)\right]^N \ket{0}.
\end{align}
where the dynamics are given by the evolution operator $\hat{\Pi}S(\Gamma_2)U(\phi)S(\Gamma_1)$ and  $\hat{\Pi}$ represents the partial measurement in one mode.

Analytically, using the relation eq.\,\eqref{SUS}, 
\begin{align}
S(r e^{i\theta})^{\dag}\hat{a}S(r e^{i\theta}) = \cosh(r)\hat{a} - e^{i\theta} \sinh(r) \hat{a}^{\dag}
\end{align}
and 
\begin{align}
e^{i\hat{a}^{\dag} \zeta \hat{a}} \hat{a} e^{-i\hat{a}^{\dag} \zeta \hat{a}} = e^{-i\zeta} \hat{a},
\end{align}
the dynamics given by the eq.\,\eqref{eqa1} can be simplified in case when $\hat{\Pi} = \hat{a}$ and $\Gamma_{1(2)} = r$ to,
\begin{align}
W_a = U^{N}(\phi) [X(r,\phi)\hat{a} + Y(r,\phi)\hat{a}^{\dag}] S(\Gamma_f (\phi))
\end{align}
where $\Gamma_f (\phi)$ is same as it is in eq.\,\eqref{simple_loopeq}, and $X$ and $Y$ are complicated functions of $\phi$ and $r$ that change with each loop.
For numerical analysis, the mapping to $(N+1)$ modes where $N$ is the number of loops, as shown in the Fig.\,\ref{SU11-b}-(b), is used. 
The photon intensity for this scheme is shown in Fig.\,\ref{Int_adaptive}, with respect to the unknown phase $\phi$ and number of loops $N$. The measured mode of the SU(1,1)-interferometer is reinitialized to vacuum state. It can be seen that the photon intensity in the partial feedback scheme increases with the number of loops $N$ and decreases with the phase $\phi$.

\subsection{SU(1,1)-interferometer in the phase-space formalism }

Any Gaussian state $\rho$ can be defined by the characteristic function of the Gaussian form\,\cite{SL15},
\begin{align}
\chi_{\rho}(\xi) &= Tr\left[\rho D(\xi)\right] \nonumber \\
&= \exp\left[-i d^{\dag} K \xi - \frac{1}{4} \xi^{\dag} \sigma \xi  \right]
\end{align} 
where, $D(\xi) = \exp(\hat{A}^{\dag} K \xi)$ is the Weyl displacement operator 
and they can be described by their first and second moments, known as displacement vector $d$ and covariance matrix $\sigma$, respectively,\,\cite{GC02} such that
\begin{align}
d_{i} &= Tr \left[\rho \hat{A}_{i} \right] \nonumber  \\
\sigma_{ij} &= Tr\left[\rho \left\{ \Delta\hat{A}_{i}, \Delta\hat{A}_{j} \right\} \right]
\end{align}
where $\hat{A} = (\hat{a}_1, \hat{a}^{\dag}_{1}...\hat{a}_{j},\hat{a}^{\dag}_{j}, ... )^{T}$ is the vector of creation and annihilation operators, $\Delta\hat{A} = \hat{A} - d$ and $\{.,.\}$ is the anti-commutation operator.
Since the input and the output states of SU(1,1)-interferometer are only squeezed states then $d=0$ throughout this work and the states are described by solely the covariance matrix $\sigma$.

The covariance matrix of a single-mode vacuum state is $\mathcal{I}_2 /2 $ where $\mathcal{I}_2$ is the two-dimensional identity matrix. Therefore, the input state in the phase-space is given by $\sigma_{in} = (\mathcal{I}_2 \oplus \mathcal{I}_2)/2$. The two-mode squeezing operator $S(\Gamma)$ is defined by,
\begin{align}
\tilde S(\Gamma) = \begin{pmatrix}
A & C \\
C^{T} & B
\end{pmatrix}
\end{align}
where,
\begin{align*}
A = B = \begin{pmatrix}
\cosh(r) & 0 \\
0 & \cosh(r)
\end{pmatrix}.
\end{align*}
and 
\begin{align*}
C = \begin{pmatrix}
0 & -e^{i\theta}\sinh(r) \\
-e^{-i\theta} \sinh(r) & 0
\end{pmatrix}
\end{align*}
and $C^{T} $ is the transpose of $C$ and $\Gamma = re^{i\theta}$.
The phase operator $U(\phi)$ consists of the relative phase between the two modes and is given by,
\begin{align}
U(\phi) = \begin{pmatrix}
\Phi & 0 \\
0 & \mathcal{I}
\end{pmatrix}
\end{align}
where
\begin{align} 
\Phi = \begin{pmatrix}
\exp(i\phi) & 0 \\
0 & \exp(-i \phi)
\end{pmatrix}.
\end{align}

Substituting the above matrices into eq.\,\eqref{SUevo} will give the covariance matrix $\sigma_{f}$, for the standard SU(1,1) interferometer,
\begin{align}\label{cov_s}
\sigma_{f} &= S(\Gamma)U(\phi)S(\Gamma)\sigma_{in}S(\Gamma)^{\dag}U(\phi)^{\dag}S(\Gamma)^{\dag} \nonumber \\
&=W \sigma_{in} W^{\dag}.
\end{align}

Similarly, the covariance matrix of the sequential feedback scheme in the SU(1,1) interferometer after $N$-loops is,
\begin{align}
\sigma_s &= W_s \sigma_{in} W_{s}^{\dag}
\end{align}
where $W_s = \left( S(\Gamma)U(\phi)S(\Gamma) \right)^{N} \equiv W^{N}$,
and the covariance matrix of the partial feedback scheme after $N$-loops is
\begin{align}
\sigma_a &= W_a \sigma_{in} W_{a}^{\dag} 
\end{align}
where $W_a = \left( \prod_{j=2}^{N+1} S_{1,j}(\Gamma) U(\phi) S_{1,j}(\Gamma) \right)$ which is the unitary constructed from Fig.\,\ref{SU11-b}-(b). $S_{1,j}(\Gamma)$ is a $(N+1)$-dimensional matrix with two-mode squeezing operator acting between the modes $(1,j)$ which are active and identity at the rest of the modes as illustrated in Fig.\,\ref{SU11-b}-(b).

\begin{figure}
\centering
\includegraphics[width=0.5\textwidth]{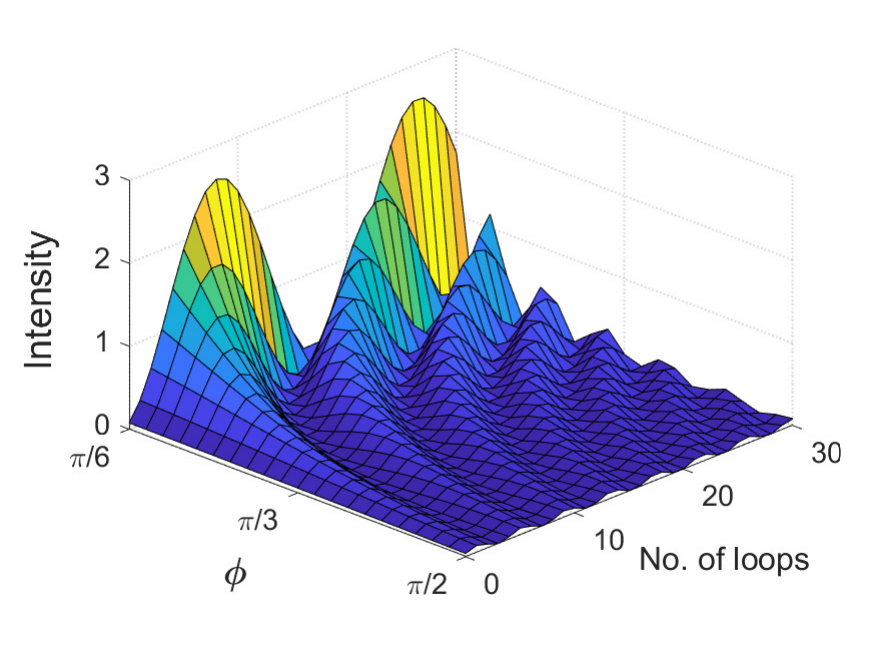}
\caption{Photon intensity in the SU(1,1) interferometer in the sequential feedback scheme, with respect to the number of loops, and phase $\phi$, for squeezing parameter $r=0.1$. The input states are the vacuum states. }
\label{Int_sequential}
\end{figure}

\begin{figure}
\centering
\includegraphics[width=0.5\textwidth]{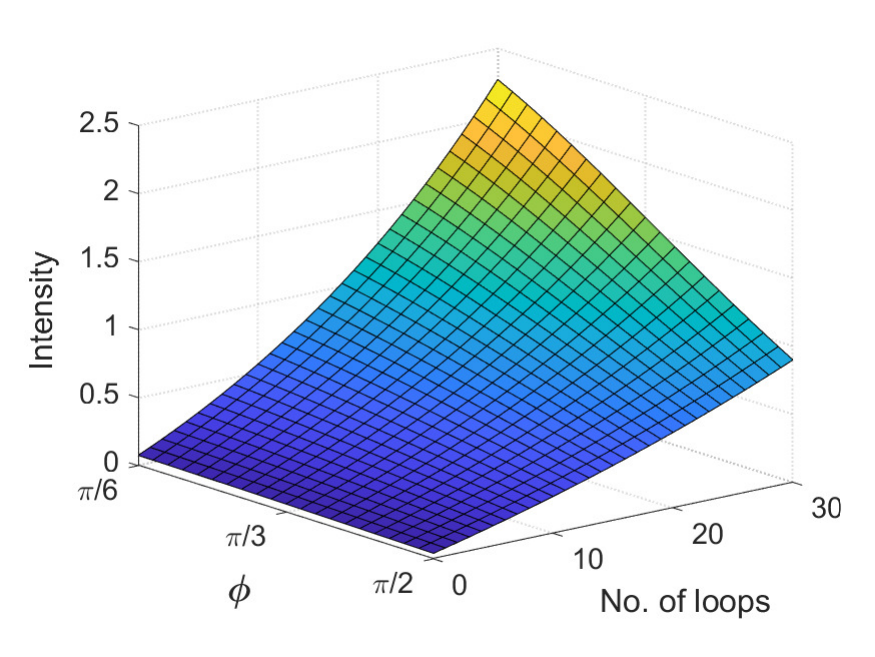}
\caption{Photon intensity in the SU(1,1) interferometer in partial feedback scheme, with respect to the number of loops, and phase $\phi$, for squeezing parameter $r=0.1$. The modes are initialized to the vacuum state.}
\label{Int_adaptive}
\end{figure}


The intensity in each mode of SU(1,1) interferometer can also be determined from the covariance matrix $\sigma$, which are simply the diagonal elements of $(\sigma - \mathcal{I}/2)$. Intensities in Fig.\,\ref{Int_sequential} and \ref{Int_adaptive} can be reconstructed using this formalism for both the feedback schemes.

\section{Phase estimation in SU(1,1) interferometer with feedback \label{secQFI}}

The maximum amount of information that can be gained about an unknown phase in an unbiased estimator can be quantified using Quantum Fisher information (QFI). According to the quantum Cramer-Rao bound, the variance of the unknown parameter $\phi$ using any unbiased estimator is lower bounded by the QFI and is given by,
\begin{align}\label{QF}
(\Delta\phi)^{2} \geq \frac{1}{MH(\phi)}
\end{align}
where $M$ is the number of independent experiments and $H(\phi)$ is the QFI of the probe state.

\subsection{QFI in an ideal SU(1,1)-interferometer with feedback}
As mentioned in the previous section, the output of the SU(1,1)-interferometer is a two-mode squeezed state $\ket{\Psi(\phi)}$ and is fully characterized by its covariance matrix $\sigma$. 
The QFI of an unknown phase $\phi$ in a pure Gaussian state $\rho$ is given by,\,\cite{J14,M13},
\begin{align}\label{H}
H(\phi) = \frac{1}{4} Tr\left[\left(\sigma^{-1} \partial_{\phi}\sigma \right)^2\right] + \partial_{\phi}d^{\dag} \sigma^{-1} \partial_{\phi}d,
\end{align}
and QFI in a noisy Gaussian state $\rho_{th}$ can be approximated as\,\cite{J14},
\begin{align}\label{Hth}
H_{th}(\phi) = \frac{1}{2} Tr\left[\left(\sigma^{-1}_{th} \partial_{\phi}\sigma_{th} \right)^2\right] + \partial_{\phi}d^{\dag}_{th} \sigma_{th}^{-1} \partial_{\phi}d_{th}.
\end{align}
where $\partial_{\phi}$ is the derivative with respect to $\phi$, $\sigma$ ($\sigma_{th}$) is the covariance matrix of the pure (noisy) Gaussian state and $d$ ($d_{th}$)is the displacement vector of pure (noisy) state\,\cite{J14, M13}.
For a pure squeezed state, the displacement vector $d$ is zero and thus the QFI reduces to,
\begin{align}\label{HSQ}
H(\phi) = \frac{1}{4} Tr\left[\left(\sigma^{-1} \partial_{\phi}\sigma \right)^2\right].
\end{align}
%

Since the probe state of SU(1,1)-interferometer is a pure squeezed state, the QFI for the unknown phase can be calculated using Eq.\,\eqref{HSQ}.
The covariance matrix $\sigma_f$ of a standard SU(1,1)-interferometer is given by Eq.\,\eqref{cov_s}.
The analytical value of the QFI in standard SU(1,1)-interferometer is $H = \sinh^2(2r)$ which is independent of the unknown phase $\phi$.
The mean photon number that probes the unknown phase in a standard SU(1,1)-interferometer is $\bar{n} = 2\sinh^2(r)$, as shown in  section\,\ref{SU11_sNa}. Thus the QFI in terms of the mean photon number is $H = \bar{n}(\bar{n}+2)$, which gives a the lower bound on the variance of $\phi$ in the Heisenberg limit.  
 

\begin{figure}
\centering
\includegraphics[width=0.5\textwidth]{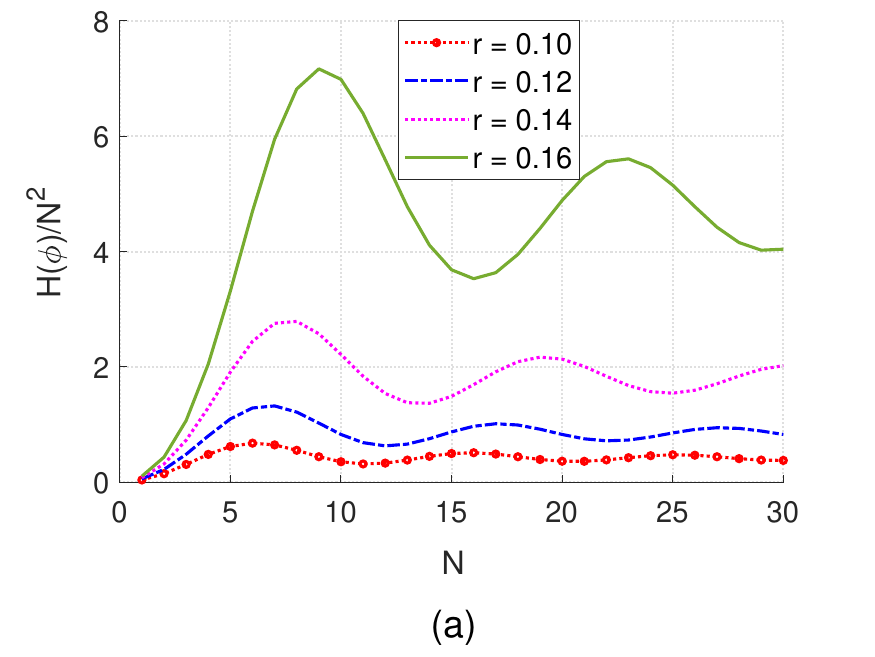}
\includegraphics[width=0.5\textwidth]{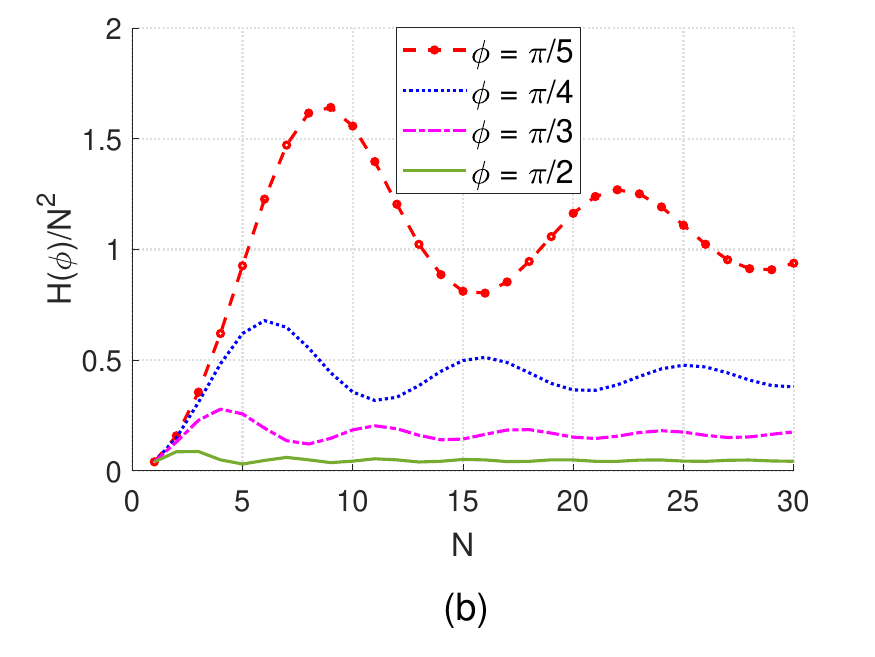}
\caption{QFI scaled by $N^2$ as a function of $N$, where $N$ is the number of loops in SU(1,1) interferometer in sequential feedback scheme (a) for different squeezing parameter $r$ when $\phi = \pi/4$ and (b) for different value of $\phi$ when $r=0.1$, respectively.}
\label{SU11_2}
\end{figure}

\begin{figure}
\centering
\includegraphics[width=0.5\textwidth]{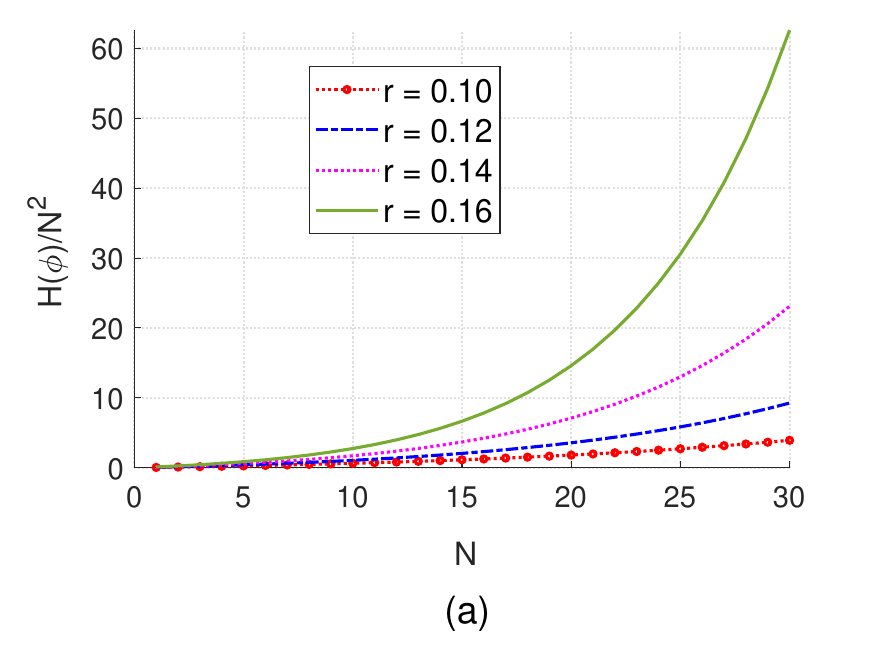}
\includegraphics[width=0.5\textwidth]{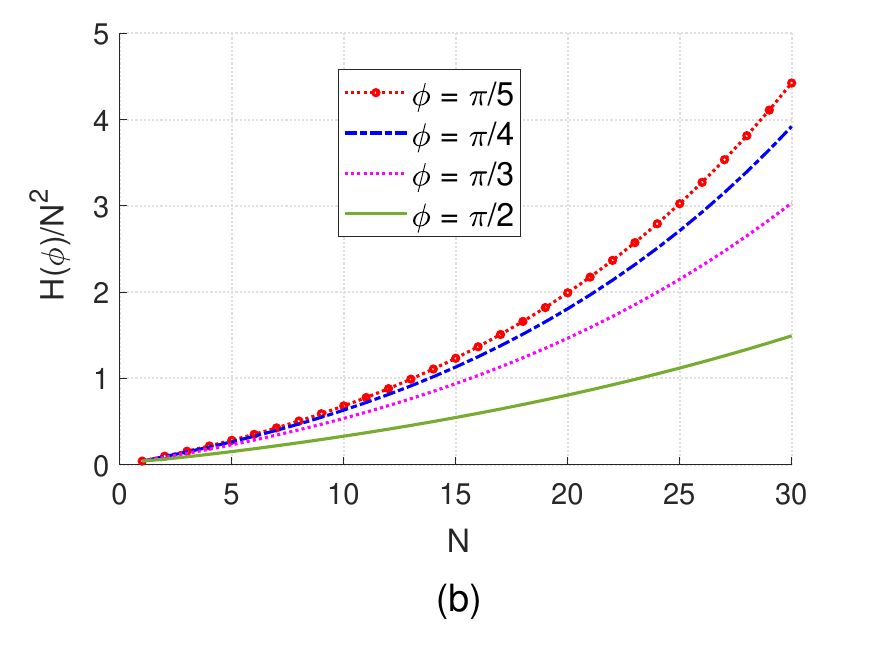}
\caption{QFI scaled by $N^2$ as a function of $N$, where $N$ is the number of loops in SU(1,1) interferometer in partial feedback scheme (a) for different squeezing parameter $r$ when $\phi = \pi/4$ and (b) for different value of $\phi$ when $r=0.1$, respectively.}
\label{SU11_1}
\end{figure}


We have analysed the QFI for single loop  i.e. two iterations in eq.\,\eqref{eqa2} for the sequential, and one iteration in eq.\,\eqref{eqa1} for the partial feedback schemes. 
Even for the simplest case of a single-loop, the QFI increases significantly and is phase $\phi$ dependent.%

\subsubsection{QFI in the sequential feedback scheme -}
The dynamics of two iteration ( equivalent to single loop) in an SU(1,1) interferometer in this feedback scheme is given by eq.\,\eqref{eqa2} where $N=2$. The QFI in this case will be,
\begin{align}
H_s(\phi) &= 4\sinh^2(2r) \left[ \cosh^2(2r) + \cosh^2(2r)\sinh^2(2r)\cos^2\phi \right. \nonumber \\
& \left. + \cos\phi\cosh(4r)\cosh^2(2r) + \cosh^2(4r)/4 \right] \nonumber \\
&= \bar{n}(2+\bar{n})\left[ 4(1+\bar{n})^4  + \bar{n}(1+\bar{n})^2(2+\bar{n})\cos^2\phi \right. \nonumber \\
&\left. + 4(1+\bar{n})^2(1+2\bar{n}(2+\bar{n}))\cos \phi + 1\right] \nonumber \\
& \approx O(\bar{n}^6),
\end{align}
where $\bar{n}$ is the mean photon number in standard SU(1,1)-interferometer. It can be seen that the QFI is of order $O(\bar{n}^6)$, which is credited to the fact that the mean photon number that probes the phase in this scheme increases with the number of loops.
The QFI in this scheme is given by eq.\,\eqref{HSQ}, where
\begin{align}
\partial_{\phi}\sigma_s &= NU^{N-1}\partial_{\phi}U S(\Gamma_f) + U^N \partial_{\phi}S(\Gamma_f) \nonumber \\
&= (N A_s + B_s) 
\end{align} 
such that the contribution of multiple interactions with the phase shifter comes from the factor $A_s = U^{N-1}\partial_{\phi}U S(\Gamma_f)$ which is the derivative of the phase operator and $B_s =  U^N \partial_{\phi}S(\Gamma_f)$ takes into account the effect of the change in the final squeezing operator with respect to phase $\phi$. Thus the QFI in this scheme will have three terms
\begin{align}\label{QFI_seq1}
H_s(\phi) &= N^2 Tr((A_s\sigma_{s}^{-1})^{2}) + 2N Tr(A_s \sigma_{s}B_s\sigma_{s}) \nonumber \\
& + Tr((B_s\sigma_{s}^{-1})^{2}).
\end{align}

Numerical simulations show that the enhancement in the QFI of this scheme is dominated by the first term of eq.\,\eqref{QFI_seq1} which is due to the multiple interactions with the phase shifter while the oscillating term is a contribution of interaction between the squeezers in the system. 
Fig.\,\ref{SU11_2} shows the scaled QFI, i.e. $H(\phi)/N^2$, with respect to $N$, where $N$ is the number of loops, for different value of $\phi$ and $r$. By scaling out the quadratic increase of the QFI with respect to number of loops, we reveal the effect of the interaction between the squeezing operation (or the nonclassicality) on the QFI in this scheme. It exhibits an oscillatory behaviour with increasing $N$, and tends to saturate around a mean scaled QFI value. This mean value is higher for the larger squeezing parameters $r$ and smaller phase values $\phi$.


%

\subsubsection{QFI in the partial feedback scheme-}
The dynamics of the single iteration (which is equivalent to single loop) in this scheme is given by Eq.\,\eqref{eqa1} for $N=1$. The QFI for this case is,
\begin{align}
H_a(\phi) &= 4\sinh^2(2r) \left[\frac{1}{2} \cosh(2r) \cosh^2(r)\left(\cos\phi +1 \right) \right. \nonumber \\
& \left. +  \frac{1}{16}\sinh^2(2r) \cosh^4(r) \left(8\cos\phi + 6 + \frac{1}{2} \cos 2\phi \right) \right] \nonumber \\
&= \bar{n}(2+\bar{n})\left[ 2+ 2(1+\bar{n})(2+\bar{n})(\cos \phi +1) \right. \nonumber \\
& \left. + \frac{\bar{n}}{16}(2+\bar{n})^3 \left(8\cos\phi + 6 + \frac{1}{2} \cos 2\phi \right) \right] \nonumber \\
& \approx O(\bar{n}^6)
\end{align} 
where $\bar{n}$ is the mean photon number in standard SU(1,1)-interferometer. 

The QFI in this scheme is given by eq.\,\eqref{HSQ}, where
\begin{align}
\partial_{\phi}\sigma_a &= NU^{N-1}\partial_{\phi}U [X(r,\phi)\hat{a} + Y(r,\phi)\hat{a}^{\dag}]S(\Gamma_f)  \nonumber \\
&+ U^N [X(r,\phi)\hat{a} + Y(r,\phi)\hat{a}^{\dag}]\partial_{\phi}S(\Gamma_f) \nonumber \\
&+ U^N \partial_{\phi}[X(r,\phi)\hat{a} + Y(r,\phi)\hat{a}^{\dag}]\partial_{\phi}S(\Gamma_f).
\end{align} 
Here, the first term also represents the contribution from the multiple phase shifters, which effectively accumulate through repeated application in the loop. 
The QFI in this scheme is expressed as, 
\begin{align}
H_a(\phi) &= Tr\left((\sigma_{a}^{-1}\partial_{\phi}\sigma_a)^2\right)
\end{align}
which results in six distinct terms.
The number of loops increases the interaction with the phase shifter, leading to a quadratic increase in QFI.
Numerical simulation shows that, unlike the sequential feedback scheme, QFI in this scheme is not solely dominated by the quadratic scaling with the number of loops. o isolate the contribution arising from the nonclassicality of the generated state, we scale the QFI as $H_a(\phi)/N^2$.
Fig.\,\ref{SU11_1} shows the scaled QFI i.e., $H_{a}(\phi)/N^2$, as a function of $N$, where $N$ is the number of loops. The plot shows a monotonic increase in the scaled QFI with $N$, indicating a non-trivial enhancement beyond simple quadratic scaling. Furthermore, the rate of increase is higher for larger values of the squeezing parameter $r$ and smaller values of the phase $\phi$.
%


\begin{figure}
\centering
\includegraphics[width=0.5\textwidth]{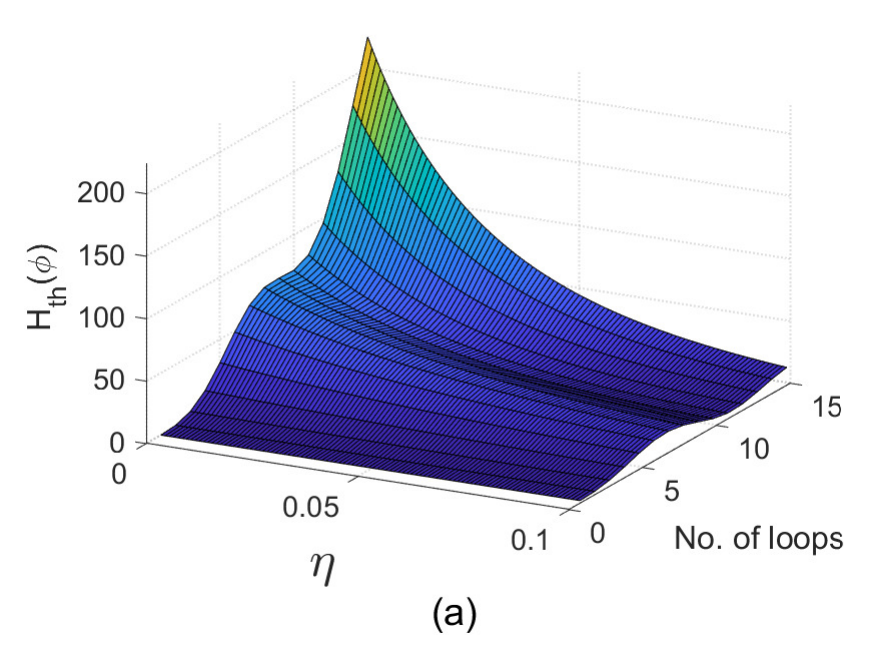}
\includegraphics[width=0.5\textwidth]{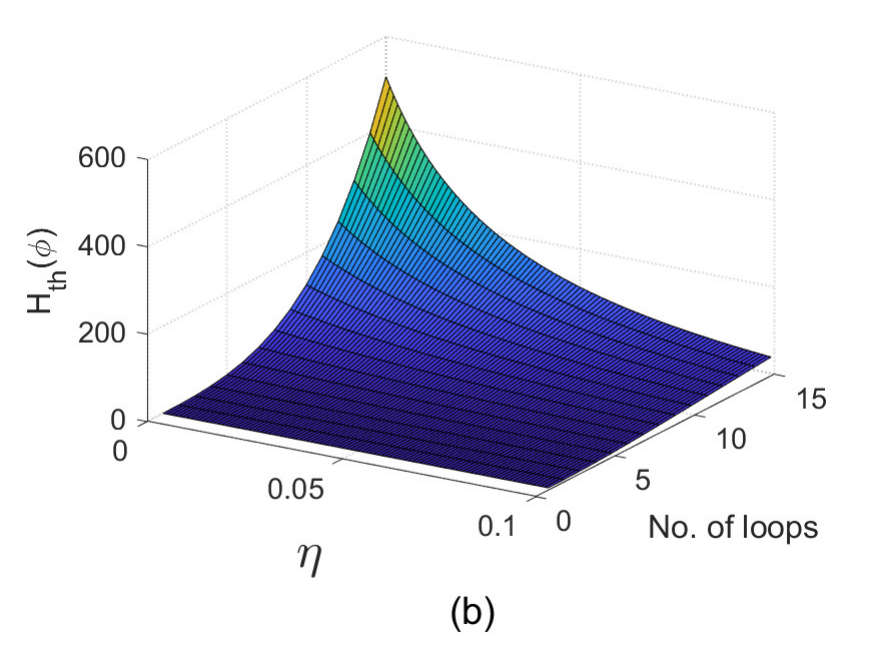}
\caption{QFI in noisy SU(1,1) interferometer $(a)$ in the sequential feedback scheme, when both the modes are in feedback loop,  and $(b)$ in the partial feedback scheme, when only single mode is in feedback loop, respectively with respect to number of loops and photon loss $\eta$. The enhancement in QFI with number of loop dies out with the increases in photon loss.}
\label{Noisy_QFI}
\end{figure}

\subsection{QFI in a noisy SU(1,1)-interferometer with feedback loop}\label{secNoisy}

The phase estimation in an SU(1,1) interferometer is sensitive to photon losses\,\cite{MT12,OH10}, which lead to a thermalized, squeezed state at the output. 
Here, we have considered the simple case of identical linear photon loss $\eta$ at both the ports of the SU(1,1)-interferometer at the end of each iteration. 

The output of the noisy SU(1,1) interferometer can be represented by a thermal state such that the covariance matrix $\sigma^{th}$ at the end of each loop is given by,
\begin{align}
\sigma^{th}_{N} = t\big(\sigma^{th}_{N-1} - \frac{\mathcal{I}}{2} \big) + \frac{\mathcal{I}}{2},
\end{align} 
where  $\eta$ is the photon loss in the system, $t = (1-\eta)$ and $\sigma^{th}_{N}$ is the covariance matrix after the iteration $N$.
The quantum Fisher information of the noisy squeezed state is given by the eq.\,\eqref{Hth} with $d = 0$ thus
\begin{align}
H_{th} = \frac{1}{2} Tr\left[\left((\sigma^{th})^{-1} \partial_{\phi}\sigma^{th} \right)^2\right].
\end{align}
Fig.\,\ref{Noisy_QFI} shows the QFI in the noisy SU(1,1) interferometer with the feedback loop and it shows that the QFI in both the feedback schemes, sequential and partial feedback, are sensitive to the photon loss in the system.
The QFI in the noisy SU(1,1) interferometer feedback schemes decreases exponentially with the photon loss $\eta$ leading to a metrological disadvantage.

\begin{figure}
\centering
\includegraphics[width=0.45\textwidth]{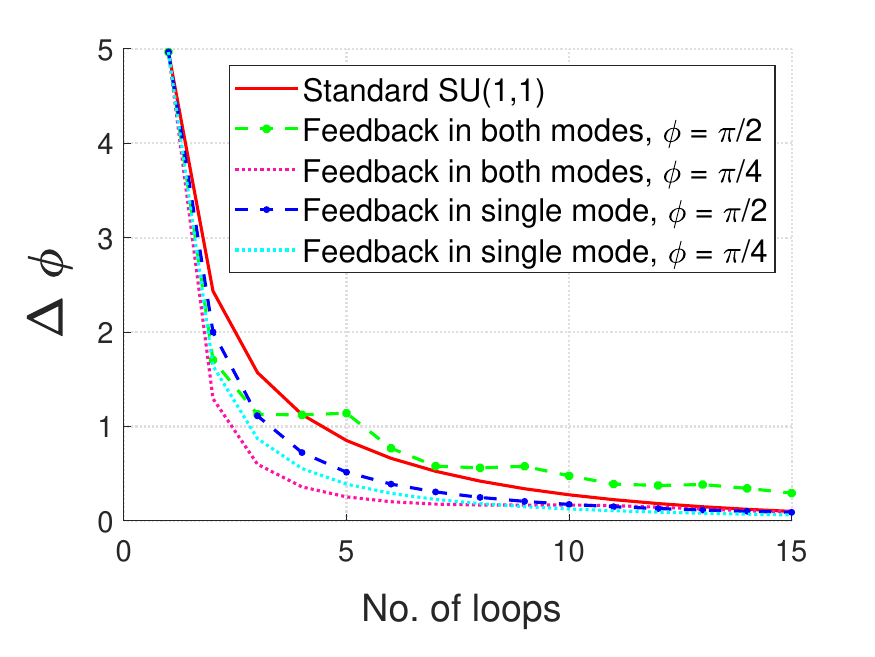}
\caption{Standard deviation in the phase estimation $\Delta \phi$ with respect to the resources which in the feedback scheme is the number of loops, and in standard SU(1,1) is the enhanced squeezing parameter $S(Nr) \equiv S(r)S(r)...S(r)$  in standard SU(1,1) interferometer for $r=0.1$. }%
\label{compa}
\end{figure}

\begin{figure}
\centering
\includegraphics[width=0.5\textwidth]{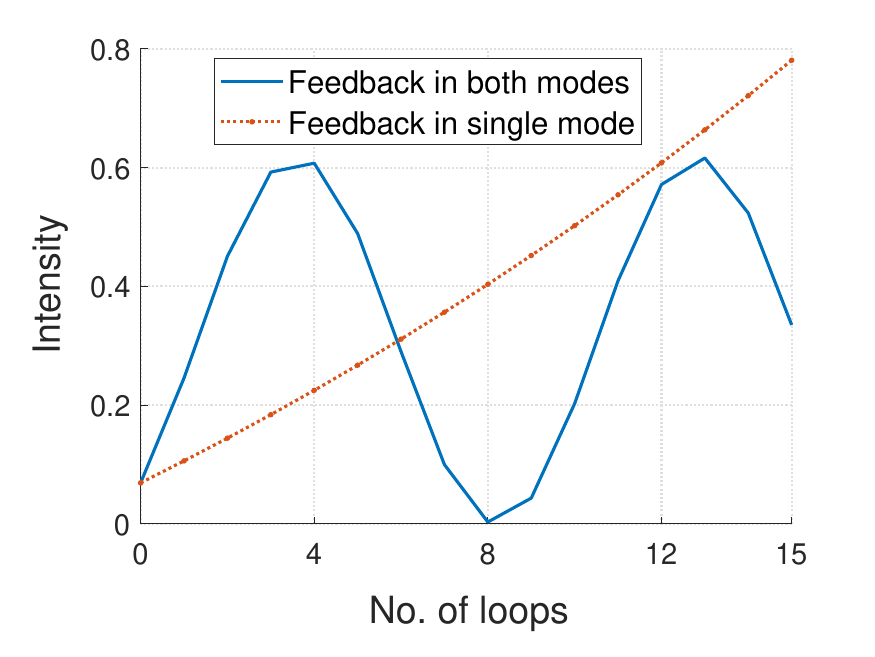}
\caption{A comparison of the mean photon number that probes the phase with respect to the number of loop for $r=0.1$ and $\phi=\pi/4$. For the initial $m$-loops the intensity in the scheme with both the modes in feedback loop is higher than the scheme with only one mode in the feedback loop. }
\label{Inten}
\end{figure}


\subsection{Phase sensitivity in SU(1,1) interferometer with feedback }\label{secSensitivity}
The phase sensitivity in the SU(1,1) interferometer for an unbiased estimator is given by the variance of the estimated phase $(\Delta\phi)^2$, in Eq.\,\eqref{QF}, which is inversely proportional to the QFI for $M=1$.  
The QFI in the standard SU(1,1) interferometer is phase independent\,\cite{QD23} while our analysis shows that QFI in both the feedback schemes depends upon the phase.

Fig.\,\ref{compa} shows a comparison of the phase sensitivity in both the feedback schemes and the standard SU(1,1) interferometer in terms of standard deviation (equal to square root of variance in the system). In terms of the resources, the number of loops $N$ in the feedback schemes, can be compared to the standard SU(1,1)-interferometer with a squeezing parameter of the order $(Nr)$ i.e., squeezing operator $S(Nr) \equiv S(r)S(r)...S(r)$ that probes the unknown phase. 
It shows that when the unknown phase in the system is small, the feedback schemes are the optimal choice but as the value of the unknown phase increases, the standard deviation in the feedback scheme becomes higher than the standard SU(1,1) interferometer. 
 The standard deviation, $\Delta \phi$, with the number of loops, in the sequential feedback scheme is larger than the partial feedback scheme when the phase value $\phi$ is large, as shown in Fig.\,\ref{compa} for phase $\phi = \pi/2$.
For a given phase $\phi$, the number of loops that minimizes the standard deviation in the sequential feedback scheme can be determined from the oscillation interval of the photon intensity. When the phase $\phi$ is small, the minimum standard deviation in the sequential scheme is lower than that in the partial feedback scheme and the standard SU(1,1) scheme, as illustrated in Fig.\,\ref{compa} for $\phi = \pi/4 $. Specifically, for $\phi = \pi/4$, the standard deviation in the sequential feedback scheme reaches a minimum at approximately $N=4$, and outperforms the partial feedback scheme for up to $N\approx 6$ loops.
Fig.\,\ref{Inten} shows a comparison of the photon intensity in the feedback schemes for $\phi = \pi/4$ and it shows that the intensity in the sequential feedback scheme is greater than the partial feedback scheme upto $N \approx 6$ and maximum at $N \approx 4$, where $N$ is the number of loops. In the next section we have devised a form of sequential feedback scheme where we show that the phase sensitivity can be improved by swapping the squeezing operators after a fixed number of loops.


%

\begin{figure}
\centering
\includegraphics[width=0.5\textwidth]{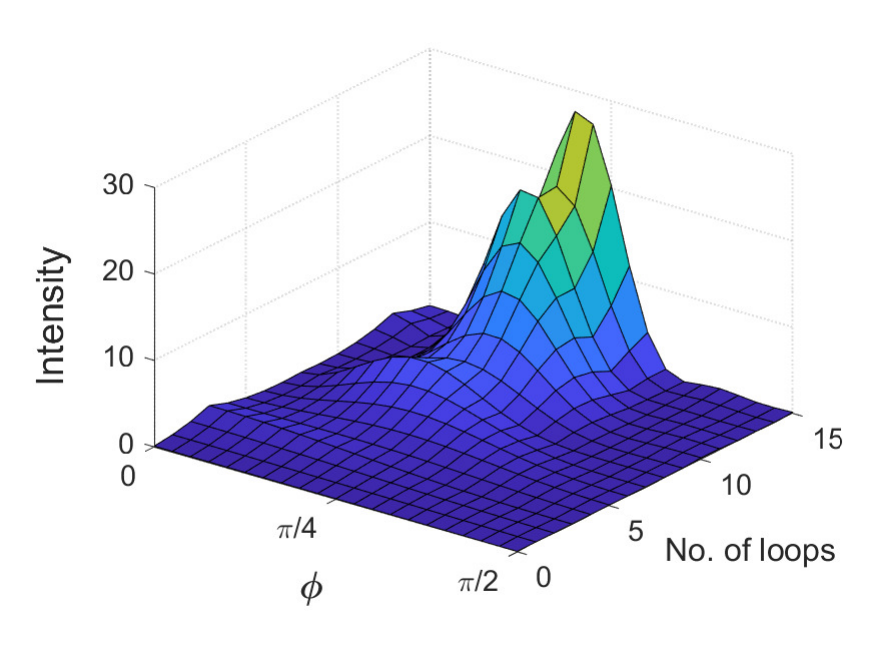}
\caption{
Photon intensity with respect to the phase $\phi$ and number of loops in the scheme with both the modes in feedback loop with swapping of squeezed state after every $4$-loops for $r=0.1$. Intensity near $\phi = \pi/4$ is increasing with the number of loops while for rest of the phases, intensity is very low (comparatively negligible). 
}
\label{sIntensity}
\end{figure}

\begin{figure}
\centering
\includegraphics[width=0.5\textwidth]{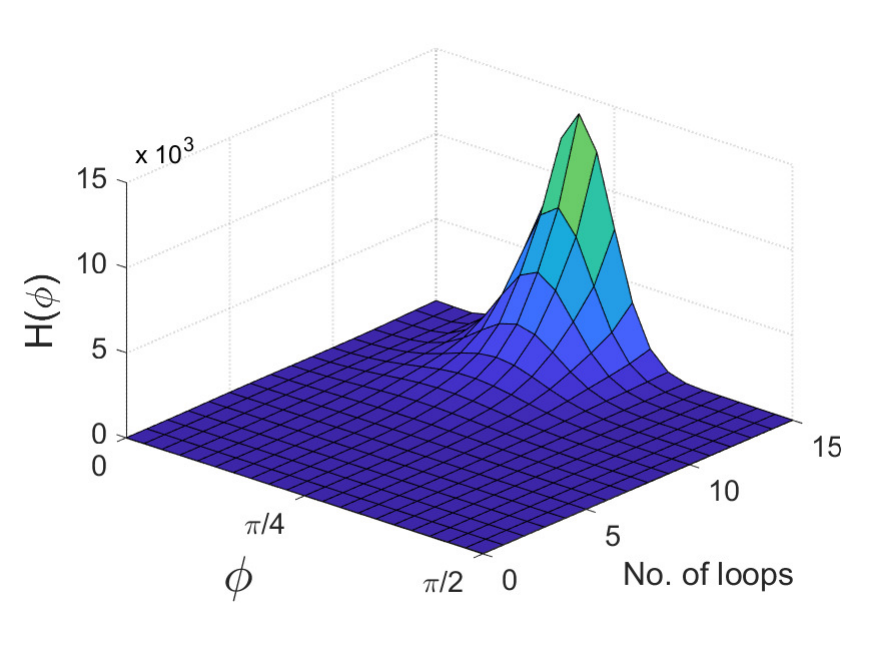}
\caption{Quantum Fisher information with respect to the phase $\phi$ and number of loops in the scheme with both the modes in feedback loop with swapping of squeezed state after every $4$-loops for $r=0.1$. The QFI in the system is sufficiently high near $\phi = \pi/4$ compared to any other feedback schemes. }
\label{sQFI}
\end{figure}


\section{Analysis of the special sequential scheme with swapping of squeezing operation \label{secSpecial}}

As shown in the previous section, the phase sensitivity of the sequential schemes reduces if the unknown phase is large, but a comparison of photon intensity and the phase sensitivity shows that interval of oscillation of the photon intensity in the sequential scheme plays a crucial role. Here, we have devised a special sequential scheme where after every $k (<N)$-loops, the squeezing operators are swapped between $S(r)$ and $S(-r)$. Swapping is performed after every $k$-loops which is approximately equal to the half of the interval of oscillation. 

In general, the output state of the sequential scheme with swapping of squeezing operator is given by,
\begin{align}\label{sevol}
\ket{\Psi(\phi)}_{ss} &= W_{ss}\ket{0} \nonumber \\
&= [S(r)U(\phi)S(r)]^{k}... \nonumber \\
&[S(-r)U(\phi)S(-r)]^{k} [S(r)U(\phi)S(r)]^{k} \ket{0}
\end{align}
where $k$ is half of the period of oscillation of the intensity in the sequential scheme for the phase $\phi$, respectively, thus it can be calculated from the intensity plot for sequential scheme shown in Fig.\,\ref{Int_sequential}.

Fig.\,\ref{Inten} shows the intensity distribution comparison for $\phi=\pi/4$ in the feedback schemes, respectively. It shows that the interval of oscillation for $\phi =\pi/4$ is $(N \approx 8)$ thus $k\approx 4$. Fig.\,\ref{sIntensity} shows the intensity distribution when the squeezing operations are swapped between $S(r)$ and $S(-r)$ after every $k=4$ feedback loops. It can be seen from the plot that intensity is maximum around $\phi=\pi/4$. 
The numerical analysis of the QFI in this special case of sequential feedback scheme is shown in Fig.\,\ref{sQFI} and it increases with the number of loops when  $\phi$ is close to $\pi/4$ when the swapping is performed after $k=4$ loops. It can be deduced that this case of sequential feedback scheme can be used to provide enhanced QFI for the estimation of small variation in the phase $\phi$. Thus a small change in the initial phase can be estimated with high efficiency and the interval of swapping $k$ can be calculated using the initial phase information. 

Another interesting observation is that the interval of oscillation is large when $\phi$ is small thus it shows that $k$- is large when the known phase difference between the modes is small. In situations like this swapping of the squeezed state is almost same as the simple feedback scheme in terms of resources therefore in an experiment the initial phase $\phi$ will play a crucial role in order to determine the most efficient feedback scheme.


\section{Conclusion \label{concl}}

We have explored the SU(1,1)-interferometer with feedback and have focused on two feedback schemes, $(a)$ when both modes of SU(1,1)-interferometer are in loop such that the output of the first SU(1,1)-interferometer is the input of second interferometer, called the sequential scheme and (b) when only one mode of the interferometer is in the loop while other arm is measured, termed as the partial scheme.
Due to the feedback, the mean photon number in the interferometer changes with each loop. Our analysis shows that the intensity distribution in the sequential scheme oscillates with the number of loops and the period of oscillation depends upon the phase $\phi$. 
The QFI in this scheme is a quadratic function of number of loops with an oscillating component that depends upon the photon number.
For the partial feedback scheme, where only one mode is in the loop and the other port is reinitialized to the vacuum state, the intensity distribution increases with the number of loops. 
The QFI increases with the number of loops and the rate at which the QFI increases in this scheme is faster than the sequential scheme. 

In both the feedback schemes, due to the complete and non-adaptive state feedback process, photons interact multiple times with the squeezing operations and phase shifters, as illustrated in Fig.\,\ref{SU11-b}. 
Our analysis shows that the QFI in these schemes arises from the cumulative effect of these repeated interactions. Specifically, the analytical analysis of the QFI shows that the repeated application of phase shifters leads to a quadratic scaling of the QFI with respect to the number of loops in both feedback schemes. In contrast, the interactions involving the squeezing operations produce an oscillatory behaviour of the QFI in the sequential feedback scheme and a monotonic increase in the partial feedback scheme. Overall, phase estimation in these feedback schemes is determined by the underlying system dynamics rather than solely by repeated phase shifts or the presence of highly nonclassical states. This highlights the uniqueness and novelty of the proposed feedback approach for quantum phase estimation.

We have also studied the effect of photon loss in both feedback schemes, where QFI decreases with photon loss in the system. 
In the loss less regime, a comparison of the standard deviation $(\Delta\phi)$ in the feedback schemes is shown that depends on the mean photon number in the SU(1,1) interferometer. The analysis shows that the sequential scheme is more advantageous when the value of phase $\phi$ and number of loop is small, but if the number of loops increases the partial feedback scheme performs better for phase estimation.

We have also discussed a scheme based on sequential SU(1,1) interferometer where around a particular phase $\phi$ one can enhance the intensity and the QFI.  This is more efficient for the cases where a small change in the phase is to be estimated near an initial phase. This scheme involves swapping the squeezing operation after a fixed number of loops $k$ which is equal to the half of the interval of oscillation of photon intensity in the sequential scheme for the initial phase information.

The model presented here has the potential to be adapted for experimental proposals, where feedback is natural in the system, such as optical delay loops, a widely used technique in photonics and quantum optics. It enables repeated interactions between the photons and the quantum system by looping the photon through the system multiple times\,\cite{WX22}. The model presented here will improve the phase sensitivity by increasing the effective interaction time.

\section*{Acknowledgment}

We acknowledge support from the European Union’s Horizon Europe research and innovation program under EPIQUE Project (Grant Agreement No. 101135288). I.J. is grateful for the financial support from GACR of the Czech Republic under Grant No. 23-07169S.


\begin{thebibliography}{30}

\bibitem{CFP17} C. L. Degen, F. Reinhard, and P. Cappellaro, Quantum sensing, 
\href{https://doi.org/10.1103/RevModPhys.89.035002}{Rev. Mod. Phys. 89, 035002, 2017}.

\bibitem{M16} Marco Genovese, Real applications of quantum imaging,
\href{https://doi.org/10.1088/2040-8978/18/7/073002}{J. Opt. 18 073002, 2016}.

\bibitem{MK93} M. J. Holland and K. Burnett,  Interferometric detection
of optical phase shifts at the heisenberg limit,
\href{https://doi.org/10.1103/PhysRevLett.71.1355}{ Phys. Rev. Lett. 71, 1355–1358, 1993}.

\bibitem{GL06} V. Giovannetti, S. Lloyd, and L. Maccone, Quantum metrology, 
\href{https://doi.org/10.1103/PhysRevLett.96.010401}{Phys. Rev. Lett. 96, 010401, 2006}.

\bibitem{DD09} U. Dorner, R. Demkowicz-Dobrzanski, B. J. Smith, J. S. Lundeen, W. Wasilewski, K. Banaszek, and I. A. Walmsley, Optimal Quantum Phase Estimation, 
\href{ https://doi.org/10.1103/PhysRevLett.102.040403}{Phys. Rev. Lett. 102, 040403, 2009}.

\bibitem{GO12} M. G. Genoni, S. Olivares, D. Brivio, S. Cialdi, D. Cipriani, A. Santamato, S. Vezzoli, and M. G. A. Paris, Optical interferometry in the presence of large phase diffusion,
\href{https://doi.org/10.1103/PhysRevA.85.043817}{Phys. Rev. A 85, 043817, 2012}.

\bibitem{DZ20} A. Datta, L. Zhang, N. Thomas-Peter, U. Dorner, B. J. Smith, and I. A. Walmsley, Quantum metrology with imperfect states and detectors,
\href{https://doi.org/10.1103/PhysRevA.83.063836}{Phys. Rev. A 83, 063836, 2020}.

\bibitem{AP97} G. M. D'Ariano and M. G. A. Paris, Arbitrary precision in multipath interferometry, 
\href{https://doi.org/10.1103/PhysRevA.55.2267}{Phys. Rev. A 55, 2267, 1997}.

\bibitem{SB03} J. Soderholm, G. Bjork, B. Hessmo, and S. Inoue, Quantum limits on phase-shift detection using multimode interferometers,
\href{https://doi.org/10.1103/PhysRevA.67.053803}{Phys. Rev. A 67, 053803, 2003}.


\bibitem{AA10} I. Afek, O. Ambar, and Y. Silberberg, High-NOON States by Mixing Quantum and Classical Light,
\href{https://doi.org/10.1126/science.1188172}{Science, 328, 5980, pp. 879-881, 2010}.

\bibitem{NO07} T. Nagata, R. Okamoto, J. L. O'Brien, K. Sasaki, and S. Takeuchi, Beating the Standard Quantum Limit with Four-Entangled Photons,
\href{https://doi.org/10.1126/science.1138007}{Science, 316, 5825, pp. 726-729, 2007}.

\bibitem{AR10}  P. M. Anisimov, G. M. Raterman, A. Chiruvelli, W. N. Plick, S. D. Huver, H. Lee, and J. P. Dowling, Quantum metrology with two-mode squeezed vacuum: Parity detection beats the Heisenberg limit,
\href{https://doi.org/10.1103/PhysRevLett.104.103602}{ Phys. Rev. Lett. 104, 103602, 2010}.

\bibitem{D08}  J. P. Dowling, Quantum optical metrology – the lowdown on high-N00N states,
\href{https://doi.org/10.1080/00107510802091298}{Contemp. Phys. 49, 125–143, 2008.}

\bibitem{JM11} J. Joo, W. J. Munro, and T. P. Spiller, Quantum metrology with entangled coherent states,
\href{https://doi.org/10.1103/PhysRevLett.107.083601}{Phys. Rev. Lett. 107, 083601,  2011.}

\bibitem{CGB03}  R. A. Campos, C. C. Gerry, and A. Benmoussa, Optical interferometry at the heisenberg limit with twin fock states and parity measurements,
\href{https://doi.org/10.1103/PhysRevA.68.023810}{ Phys. Rev. A 68, 023810, 2003.}

\bibitem{YM96} B. Yurke, S. L. McCall, and J. R. Klauder, SU(2) and SU(1,1) interferometers,
\href{https://doi.org/10.1103/PhysRevA.33.4033}{Phys. Rev. A 33, 4033, 1996}


\bibitem{AG17} B. E. Anderson, P. Gupta, B. L. Schmittberger, T. Horrom, C. Hermann-Avigliano, K. M. Jones, and P. D. Lett, Phase sensing beyond the standard quantum limit with a variation on the SU(1,1) interferometer,
\href{https://doi.org/10.1364/OPTICA.4.000752}{Optica 4, 7, pp. 752-756, 2017}.

\bibitem{LZ23}  M Liu, L. Zhang and H Miao, Adaptive protocols for SU(1,1) interferometers to achieve ab initio phase estimation at the Heisenberg limit, 
\href{https://doi.org/10.1088/1367-2630/ad042f}{New J. Phys. 25 103051, 2023}.

\bibitem{C20} C. M. Caves, Reframing SU(1,1) interferometry, 
\href{https://doi.org/10.1002/qute.201900138}{Adv. Quantum Technol. 3(11), 1900138 (2020).}

\bibitem{CZ22} S. K. Chang, W. Ye, H. Zhang, L. Y. Hu, J. H. Huang, and S.
Q. Liu, Improvement of phase sensitivity in an SU(1,1) interferometer via a phase shift induced by a Kerr medium,
\href{https://doi.org/10.1103/PhysRevA.105.033704}{Phys. Rev. A 105(3), 033704 (2022).}

\bibitem{SL17} S. S. Szigeti, R. J. Lewis-Swan, and S. A. Haine, Pumped-
up SU(1, 1) interferometry, 
\href{https://doi.org/10.1103/PhysRevLett.118.150401}{Phys. Rev. Lett. 118(15), 150401 (2017).}

\bibitem{DC20} W. Du, J. F. Chen, Z. Y. Ou, and W. Zhang, Quantum dense metrology by an SU(2)-in-SU(1,1) nested interferometer, 
\href{https://doi.org/10.1063/5.0012304}{Appl. Phys. Lett. 117(2), 024003 (2020).}


\bibitem{MC12} A. M. Marino, N. V. Corzo Trejo, and P. D. Lett, Effect of losses on the performance of an SU(1,1) interferometer,
\href{https://doi.org/10.1103/PhysRevA.86.023844}{Phys. Rev. A 86, 023844, 2012.}



\bibitem{JX20}J. Xin, X. Pan, X. Lu, J. Kong, G. Li, and X. Li, Entanglement Enhancement from a Two-Port Feedback Optical Parametric Amplifier,
\href{https://doi.org/10.1103/PhysRevApplied.14.024015}{Phys. Rev. Applied 14, 024015 (2020).}

\bibitem{HZ21} Y. Han, Z. Zhang, Z. Zhou, J. Qu, J. He, and J. Wang, Enhancement of squeezing with cascaded and coherent feedback-controlled degenerate optical parametric amplifiers,
\href{https://opg.optica.org/josab/abstract.cfm?URI=josab-38-10-3096}{ J. Opt. Soc. Am. B 38, 3096-3104 (2021).}

\bibitem{G24} G. Jiao, Enhanced phase sensitivity in a feedback-assisted interferometer, 
\href{https://doi.org/10.1088/1367-2630/ad69b9}{New J. Phys. 26 083005 (2024).}

\bibitem{LX21} D. Liao, J. Xin, and J. Jing, Nonlinear interferometer based on two-port feedback nondegenerate optical parametric amplification, 
\href{https://doi.org/10.1016/j.optcom.2021.127137}{Opt. Commun. 496, 127137 (2021)}.

\bibitem{BG15} D. Burgarth, V. Giovannetti, A. Kato, and K. Yuasa, Quantum estimation via sequential measurements,
\href{ https://doi.org/10.1088/1367-2630/17/11/113055}{ New J. Phys. 17 113055 (2015).}

\bibitem{M11} H. Mabuchi, Nonlinear interferometry approach to photonic sequential logic,
\href{https://doi.org/10.1063/1.3650250}{Appl. Phys. Lett. 99, 153103 (2011).}

\bibitem{W94} H. M. Wiseman, Quantum theory of continuous feedback, 
\href{ https://doi.org/10.1103/PhysRevA.49.2133}{Phys. Rev. A 49, 2133 (1994).}

\bibitem{EL20} M. Engelkemeier, L. Lorz, S. De , B. Brecht , I. Dhand, M. B. Plenio, C. Silberhorn, and J. Sperling, Quantum photonics with active feedback loops, 
\href{https://doi.org/10.1103/PhysRevA.102.023712}{Phys. Rev. A 102, 023712 (2020).}

\bibitem{ZL17} J. Zhang, Y. Liu, R. Wu , K. Jacobs, and F. Nori, Quantum feedback: Theory, experiments, and applications, 
\href{https://doi.org/10.1016/j.physrep.2017.02.003}{Physics Reports Volume 679,  Pages 1-60 (2017).}


\bibitem{CI17} C. Hamilton, S. Barkhofen, L. Sansoni, I. Jex and C. Silberhorn, Driven discrete time quantum walks,
\href{https://doi.org/10.1088/1367-2630/18/7/073008}{New J. Phys. 18 073008, 2016.} 

\bibitem{SC23} S. Singh, C. Hamilton, and I. Jex, Phase estimation in driven discrete-time quantum walks, 
\href{https://doi.org/10.1103/PhysRevA.108.042607}{Phys. Rev. A 108, 042607, 2023.}



\bibitem{H11} A. Holevo, Probabilistic and Statistical Aspects of Quantum Theory,
\href{https://doi.org/10.1007/978-88-7642-378-9}{Springer Science and Business Media,1, 2011.}


\bibitem{P09} M. G. A. Paris, Quantum estimation for quantum technology,
\href{https://doi.org/10.1142/S0219749909004839}{Int. J. Quant. Inf. 7, 125, 2009.}

\bibitem{TA14}  G. Toth and I. Apellaniz, Quantum metrology from a quantum information science perspective,
\href{https://doi.org/10.1088/1751-8113/47/42/424006}{ J. Phys. A: Math. Theor. 47, 424006, 2014}.

\bibitem{GL04} V. Giovannetti, S. Lloyd , and L. Maccone, Quantum-Enhanced Measurements: Beating the Standard Quantum Limit,
\href{https://doi.org/10.1126/science.1104149}{Science 306, 1330–6, 2004}.

\bibitem{XW89} X. Ma and W. Rhodes, Multimode squeeze operators and squeezed states, 
\href{https://doi.org/10.1103/PhysRevA.41.4625}{Phys. Rev. A 41, 4625, 1990}.

\bibitem{SL15} D. Šafránek, A. Lee and I. Fuentes, Quantum parameter estimation using multi-mode Gaussian states,
\href{https://doi.org/10.1088/1367-2630/17/7/073016}{New J. Phys. 17 073016 (2015).}

\bibitem{GC02} G. Giedke and J. Cirac, Characterization of Gaussian operations and distillation of Gaussian states,
\href{https://doi.org/10.1103/PhysRevA.66.032316}{Phys. Rev. A 66, 032316, 2002}

\bibitem{WE03} M. Wolf, J. Eisert, and M. Plenio, Entangling Power of Passive Optical Elements,
\href{https://doi.org/10.1103/PhysRevLett.90.047904}{Phys. Rev. Lett. 90, 047904, 2003.}

\bibitem{VW02} G. Vidal, and R. Werner, Computable measure of entanglement,
\href{ https://doi.org/10.1103/PhysRevA.65.032314}{Phys. Rev. A 65, 032314, 2002.}

\bibitem{LY14} D. Li, C. Yuan, Z. Ou, and W. Zhang, The phase sensitivity of an SU(1,1) interferometer with coherent and squeezed-vacuum light,
\href{https://doi.org/10.1088/1367-2630/16/7/073020}{New Journal of Physics 16 073020, 2014.}

\bibitem{J14} Z. Jiang, Quantum Fisher information for states in exponential form,
\href{https://doi.org/10.1103/PhysRevA.89.032128}{Phys. Rev. A 89, 032128, 2014.}

\bibitem{M13} A. Monras, Phase space formalism for quantum estimation of Gaussian states,
\href{https://doi.org/10.48550/arXiv.1303.3682}{arXiv:1303.3682}

\bibitem{MT12} A. Marino, N. Trejo and P. Lett, Effect of losses on the performance of an SU(1,1) interferometer,
\href{https://doi.org/10.1103/PhysRevA.86.023844}{Phys. Rev. A 86, 023844, 2012.}

\bibitem{OH10} T. Ono and H. Hofmann, Effects of photon losses on phase estimation near the Heisenberg limit using coherent light and squeezed vacuum,
\href{https://doi.org/10.1103/PhysRevA.81.033819}{Phys. Rev. A 81, 033819, 2010.}

\bibitem{QD23} J. Qin, Y. Deng, H. Zhong, L. Peng, H. Su, Y. Luo, J. Xu, D. Wu, S. Gong, H. Liu, H. Wang, M. Chen, L. Li, N. Liu, C. Lu, and J. Pan, Unconditional and Robust Quantum Metrological Advantage beyond N00N States,
\href{https://doi.org/10.1103/PhysRevLett.130.070801}{Phys. Rev. Lett. 130, 070801 (2023).}

\bibitem{WX22} L. Wang, F. Xie, Y. Zhang, M. Xiao, and F. Liu, Adaptive optical phase estimation for real-time sensing of fast-varying signals,
\href{https://doi.org/10.1038/s41598-022-26329-1}{ Sci Rep 12, 21745 (2022).} 


\end{thebibliography}
\end{document}